\documentclass[aps,pra,twocolumn,showpacs,groupedaddress]{revtex4}  % for review and submission 
\usepackage{graphicx}  % needed for figures
\usepackage{dcolumn}   % needed for some tables
\usepackage{bm}        % for math
\usepackage{amssymb}   % for math

\begin{document}
 
%\hspace{5.2in} \mbox{}

\title{Experimental comparison of Raman and RF outcouplers for high flux atom lasers}			

\author{J.\,E.\,\,Debs}
	\email{john.debs@anu.edu.au}
	\homepage{http://atomlaser.anu.edu.au}
\author{D.\,\,D\"oring}
\author{P.\,A.\,\,Altin}
\author{C.\,\,Figl}
\author{J.\,\,Dugu\'e}
\author{M.\,\,Jeppesen}
%\author{R.\,\,Poldy}
\author{J.\,T.\,\,Schultz}
\author{N.\,P.\,\,Robins}
\author{J.\,D.\,\,Close}

\affiliation{\vspace{0.1 in}Australian Centre for Quantum-Atom Optics, Department of Quantum Science, The Australian National University, Canberra, 0200, Australia\vspace{0.1 in}}

\date{\today}

\begin{abstract}
We study the properties of an atom laser beam derived from a Bose-Einstein condensate using three different outcouplers, one based on multi-state radio frequency transitions and two others based on Raman transitions capable of imparting momentum to the beam. We first summarize the differences that arise in such systems, and how they may impact on the use of an atom laser in interferometry. Experimentally, we examine the formation of a bound state in all three outcouplers, a phenomenon which limits the atom laser flux, and find that a two-state Raman outcoupler is the preferred option for high flux, low divergence atom laser beams. 
\end{abstract}
\maketitle 
\section{Introduction}
There has been significant recent interest in atom interferometers based around the coherent atomic samples known as Bose-Einstein condensates (BECs) \cite{Andrews:1997aa,Torii:2000aa,Schumm:2005aa,Wang:2005aa,Bourdel:2006aa,Le-Coq:2006aa,Cronin:2007aa,Fattori:2008aa,Gustavsson:2008aa}, particularly due to the possibility of using squeezing to enhance the interferometric sensitivity \cite{Orzel:2001aa,Jo:2007aa,Li:2007aa,Esteve:2008aa}.  Atoms interact strongly with their environment, and this sensitivity to, for example, inertial and electro-magnetic forces makes atom interferometers a promising choice for applications as ultra-sensitive detectors of magnetic, optical, and inertial effects. Such applications could include gradiometer-based mineral sensing, and recent work has even proposed atom interferometry systems capable of testing predictions of general relativity \cite{Dimopoulos:2007aa} and detection of gravitational waves \cite{Dimopoulos:2008aa}. Indeed, atom interferometers based on the Sagnac effect, and either thermal beam or magneto-optic trap (MOT) sources, have already proven to be a competitive tool for high-precision rotation sensing \cite{gustavson:2000aa,Muller:2007aa,Muller:2008ab}. State of the art devices have reached short term sensitivities of $6\times 10^{-10}$\,rad/s/$\sqrt{\text{Hz}}$, an order of magnitude better than the best optical ring laser gyroscope \cite{gustavson:2000aa}. 

The majority of experimental work to date using Bose-condensed samples is based on {\em trapped} atomic ensembles for both interferometry and squeezing. Such trapped samples have the advantage of being easily addressed for (in principle) long interrogation times, as well as precise control over atomic trajectories by using matter-waveguides. However any trapped sample is inherently strongly coupled to the environment, which can cause substantial technical noise \cite{fnote1}. They also suffer from undesirable mean-field effects, such as dephasing, due to their high density. These effects require negating (by using, for example, a feshbach resonance) in order to take advantage of long interrogation times. The freely propagating matter waves derived from a BEC, known as atom lasers, allow the construction of a free space interferometer \cite{Doring:2008ab}, well suited to the detection of inertial effects. Their lower density ensures substantial reduction of undesirable mean-field effects, and a reasonable hypothesis is that free evolution results in significantly less technical noise arising from coupling to the environment. Their free evolution can however add additional complexities in addressing the atoms, and also limits their maximum interrogation time for typical setups. Although it remains an open question, atom lasers could potentially be more useful over trapped samples in practical applications, such as inertial sensing. 

When considering inertial sensors, current devices using thermal beam and MOT sources have the advantage of higher flux and typically shorter duty cycles (continuous in the case of a thermal beam). However, atom lasers present several attractive properties. For example, even the coldest thermal sources have a large transverse velocity width compared to atom lasers of a similar flux. Even after careful velocity selection, transverse widths can be more than 2 orders of magnitude larger (more than 4 orders in energy) \cite{Gustavson:2000ab}. This limits the efficiency of the Raman transitions used to perform the beam splitting process \cite{Kasevich:1991aa}, and thus the number of atoms contributing to the interference signal. Furthermore, the average longitudinal velocity of thermal sources results in a rather small beam separation after beam splitting for typical devices, limiting the area enclosed by the interferometer, and thus the accumulated phase shift. Atom lasers typically have a much smaller longitudinal velocity, allowing access to an enclosed area at least an order of magnitude larger for a setup of equivalent length. Alternatively, one may obtain an equivalent area with a much more compact setup using an atom laser \cite{fnote2}. Finally, second order correlations \cite{Ottl:2005aa} also give atom lasers the potential to be substantially quadrature squeezed with the aid of schemes such as those proposed in \cite{Haine:2005aa,Haine:2006aa,Bradley:2007aa,Haine:2008aa}. This could further increase the sensitivity of a shot-noise limited atom interferometer.  

Whether atom lasers offer substantial improvements in practice over thermal-based sources remains an open question.  In order to fully explore their potential, one first places stringent requirements on the beam properties, and hence the outcoupler used to achieve this. As with optical interferometry, we desire high flux, low divergence and a simple spatial mode. Since their first demonstration in 1997 \cite{Mewes:1997aa}, radio frequency (rf) outcouplers have been extensively studied in the literature, in the context of the divergence \cite{Le-Coq:2001aa,Jeppesen:2008aa}, flux \cite{Robins:2005aa,Robins:2006aa,Dugue:2007aa,Couvert:2008aa}, and transverse structure (spatial modes) of the atom laser beam \cite{Kohl:2005aa,Riou:2006aa,Dall:2007aa,Jeppesen:2008aa,Couvert:2008aa}. However, in recent years, outcouplers based on Raman transitions \cite{Hagley:1999aa,Robins:2006aa,Debs:2009aa} have been experimentally shown to reduce the divergence and improve the spatial mode when compared with rf systems operating at maximum flux \cite{Jeppesen:2008aa}. This is due to the momentum that Raman transitions are capable of imparting to the outcoupled atoms. Furthermore, in \cite{Dugue:2007aa} it was shown theoretically that a two-state coupling scheme maximizes the flux and helps simplify the longitudinal spatial mode compared to multi-state outcouplers. Our recent work on a two-state Raman outcoupler has shown that such a device is now the superior choice for satisfying the above requirements \cite{Debs:2009aa}. However, one hypothesis remains experimentally untested. 

In our previous work \cite{Robins:2006aa}, it was proposed that Raman-based outcouplers would also have the advantage of boosting the maximum flux limit when compared with an rf or any zero momentum transfer outcoupler. This limit occurs due to the presence of a bound state at high coupling strength, that shuts down operation of an atom laser produced from a magnetically confined BEC using an internal-state-changing outcoupler \cite{Robins:2005aa}. In this paper, we make an experimental comparison of atom laser shutdown for a three-state rf outcoupler and two types of Raman outcouplers. One of these Raman outcouplers is based on coupling of Zeeman levels of a single hyperfine state, forming a three-state system, and the other on coupling of different hyperfine levels, forming a two-state system. We verify our earlier predictions and find that the two-state Raman outcoupler is the desirable choice for producing high brightness atom lasers for applications of atom interferometry.

\section{Shutdown of an atom laser}

\begin{figure}[t]
\includegraphics{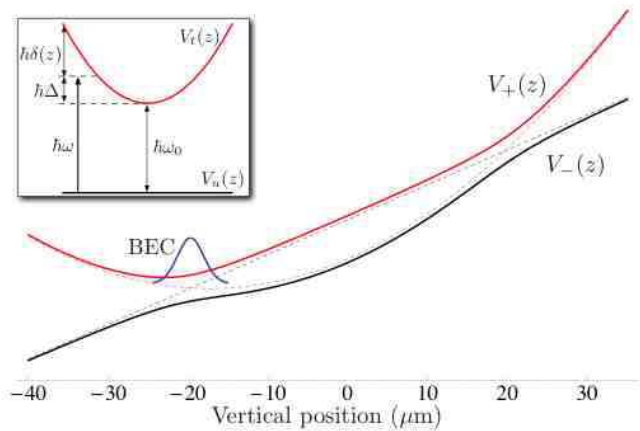}
\caption{\label{potentials} (Colour online) Dressed potentials (solid lines) formed by coupling the internal states of a two-level atom with radiation of frequency $\omega$. The inset shows the bare atomic state energies as a function of vertical position in the trap. $V_+(z)$ admits bound states whereas $V_-(z)$ does not. The dashed lines represent the potentials as the coupling is switched off, showing two crossing regions. The crossing at $z\sim-20\,\mu m$ represents the centre-of-mass position of the condensate, with the blue curve representing the BEC wavefunction and extent in the potentials. The gravitational potential is added only to the dressed state energies for reasons outlined in the main text.}
\end{figure}

\noindent The physics of atom laser shutdown is described well by the formation of dressed states and their associated potentials---a well known phenomenon of electromagnetically coupled atomic systems \cite{Cohen-Tannoudji:1977aa}. We present a simplified one-dimensional, semiclassical, two-level model to demonstrate the properties of this system. In what follows, we consider only the strong coupling limit, defined as the domain in which the kinetic term of the Hamiltonian can be neglected. A rigorous treatment of BECs in dressed potentials can be found in \cite{Zobay:2004aa}, and such potentials have been studied experimentally in the context of forming rf-induced trapping potentials for BEC \cite{Colombe:2004aa,White:2006aa,Lesanovsky:2006aa,Hofferberth:2006aa}. 

\begin{figure*}[ht]
\includegraphics[scale=0.61]{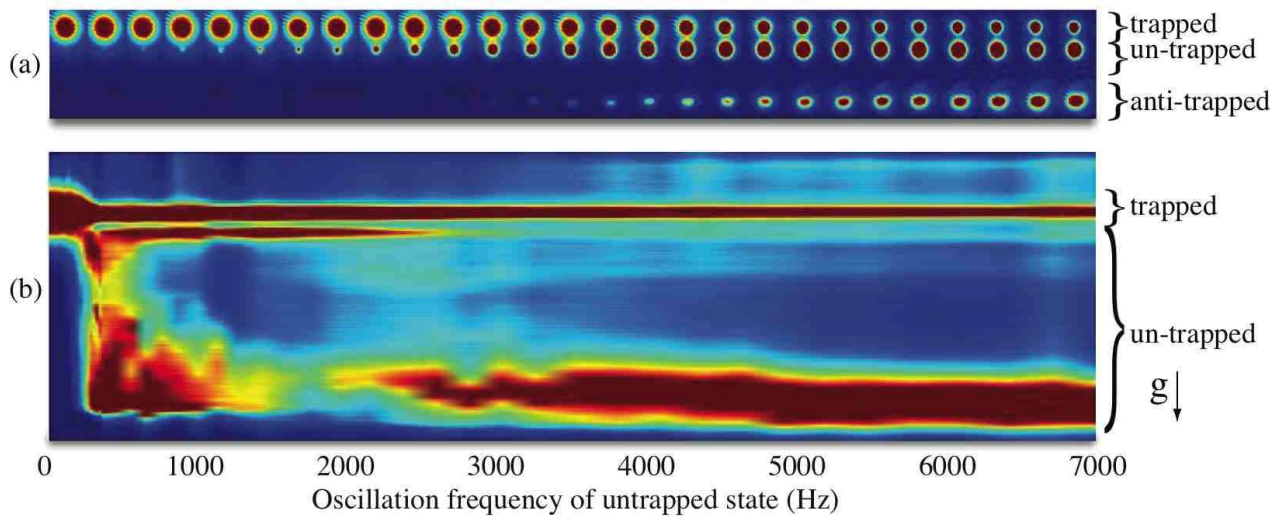}
\caption{\label{Z-Raman_abs} (Colour online) Absorption image data for a Raman outcoupler operating between Zeeman states of the $F=1$ ground state of $^{87}$Rb. (a) Images displayed as a function of the oscillation frequency of the untrapped state population, for a short (100\,$\mu$s) pulse of outcoupling. These data were used to calibrate the Rabi frequency for our setup, as described in Section III. (b) An integration of data taken for the continuous outcoupling regime. The absorption image of each individual run of the experiment has been integrated in the direction perpendicular to propagation of the atom laser beam, and each is plotted as a function of the oscillation frequency of the untrapped state population. Atom laser shutdown can be seen in the form of bound states as the coupling strength is increased.}
\end{figure*}

\begin{figure}[t]
\includegraphics[scale=1]{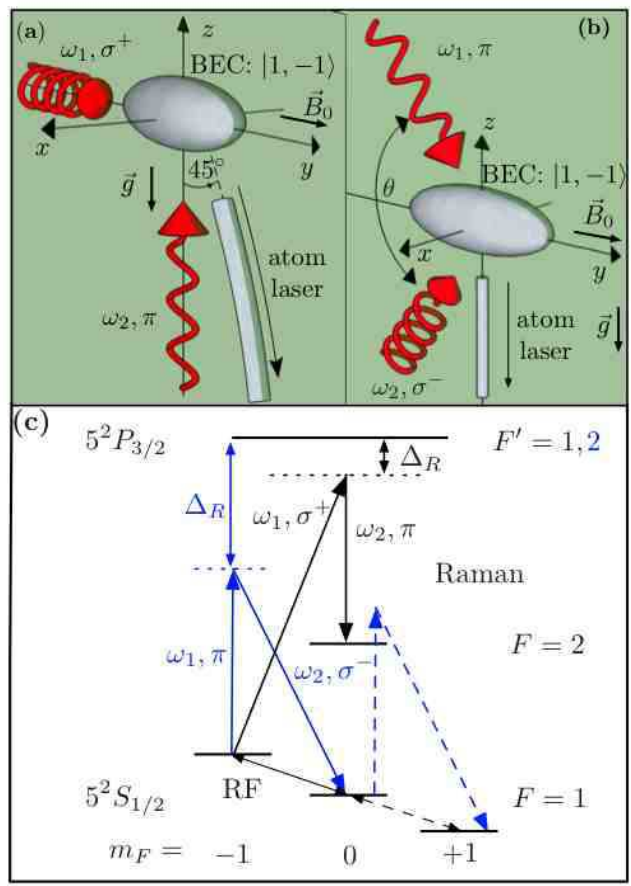}
\caption{\label{Levels} (Colour online) (a) Orientation of the Raman beams with respect to our magnetic trap for two-state hyperfine coupling. (b) Orientation of the Raman beams with respect to our magnetic trap for three-state Zeeman coupling. (c) Simplified level diagram of $^{87}$Rb for the different outcoupling schemes. An rf antenna (not shown) drives transitions between Zeeman states in the $F=1$ ground state. Optical Raman beams drive a two-photon transition between hyperfine (black) or Zeeman (blue) ground states of $^{87}$Rb.  For the Zeeman outcouplers, atoms coupled to $m_F=0$ can then also be coupled to $m_F=1$. Hyperfine splitting of the two ground states is approximately 6.834\,GHz, and Zeeman splitting is approximately 1.34\,MHz for our trap bias field. The one-photon detuning $\Delta_R$ is 300\,GHz and 90\,GHz for the Zeeman-Raman and hyperfine-Raman outcouplers respectively. All outcouplers are operated on resonance.}
\end{figure}

Consider a two-level atom in a harmonic magnetic field, with bare trapped and untrapped atomic states, $|t\rangle$ and $|u\rangle$. The internal energy of these states is given by $V_t(z)=\mu(1/2B''z^2+B_0)=1/2m\omega_z^2z^2+\hbar\omega_0$, and $V_u(z)=0$ respectively. Here, $m$ is the mass of the atom, $\omega_z$ is the trapping frequency in the vertical direction, and $\omega_0=\mu B_0/\hbar$ is the transition frequency at the magnetic field minimum $B_0$, for an atomic state with magnetic moment $\mu$. $B''$ is the curvature of the field about the minimum. The inset of Fig.\,\ref{potentials} shows the levels of the bare states as a function of vertical position, $z$, in the trap. One can rewrite $V_t(z)$ in terms of a classical coupling field of frequency $\omega$ as $V_t(z)=\hbar\delta(z)+\hbar\omega$, where $\delta(z)$ is a position dependent detuning relative to $\omega$. Alternatively, $\hbar\delta(z)=1/2m\omega_z^2z^2-\hbar\Delta$, where $\Delta=\omega-\omega_0$ is the detuning relative to the transition frequency at the magnetic field minimum. Gravity causes the condensate to sag to $z_c=-g/\omega_z^2$, where $g$ is the magnitude of acceleration due to gravity. Thus resonance does not occur at the magnetic field minimum, and $\Delta$ can be used as a measure of the sag for $\omega$ on resonance with the centre of the condensate. However, by neglecting the kinetic term in the Hamiltonian, we may treat each point in space separately, and gravity simply adds an equivalent energy offset to each state at each point. It can therefore be neglected to simplify the derivation, and then added to the resulting eigenvalues. The Hamiltonian in a frame rotating at $\omega$ can be written as:
\begin{equation}
\label{ }
H=\hbar\left(\begin{array}{cc}\delta(z) & \Omega \\\Omega & 0\end{array}\right)
\end{equation}
in the atomic basis, where the rotating wave approximation has been used. The angular Rabi frequency, $\Omega$, is a measure of the coupling strength, and is equivalent to the (on resonance) oscillation frequency of the population in a given state for a two-state system. In the case of rf coupling $\hbar\Omega=\langle t|\vec{\mu}\cdot\vec{B}_{\text{rf}}|u\rangle$ is the energy of the magnetic dipole coupling, where $|\vec{B}_{\text{rf}}|$ is the amplitude of the coupling field. In the case of Raman coupling and for large one-photon detuning $\Delta_R$ (see Fig.\,\ref{Levels}(b)), an effective two-level system is formed and $\hbar\Omega=\hbar \Omega_1\Omega_2/2\Delta_{\text{R}}$ is the energy of the two-photon coupling. Here, $\Omega_{1,2}$ represent the one-photon electric dipole couplings respectively, defined in the same manner as the magnetic dipole coupling above. A point-wise diagonalization of the Hamiltonian gives:
\begin{equation}
\label{ }
H_{\text{int}}=\left(\begin{array}{cc}V_+(z) & 0 \\0 & V_-(z)\end{array}\right)
\end{equation}
\noindent where $V_{\pm}=\hbar/2(\delta(z)\pm\sqrt{\delta(z)^2+4\Omega^2})-mgz$ are dressed potentials associated with the eigenstates (or dressed states) of the coupled system, $|+\rangle$ and $|-\rangle$. Gravity has been included for clarity. The dressed potentials are plotted in Fig.\,\ref{potentials} (solid lines) for outcoupling on resonance with the centre of the condensate. The dashed lines represent the uncoupled, (bare) atomic potentials in this frame, which cross at the position of the resonance. Clearly $|+\rangle$ is bound, and remains so indefinitely in the limit of infinite coupling.  

For a typical sequence used to produce an atom laser, outcoupling is switched on suddenly ($<200$\,ns), projecting the condensate onto the dressed basis. For a two-state system this is given by $|t\rangle=\frac{1}{\sqrt{2}}(|+\rangle-|-\rangle)$ at $\delta(z)=0$ (in other words, at the centre of the outcoupling region). For strong coupling, this expansion of $|t\rangle$ in terms of the dressed states is valid well beyond the extend of the cloud. Thus, for strong coupling a significant component of the wavefuction is in the $|+\rangle$ state and remains bound, whilst the $|-\rangle$ component of may leave the trap region. When the outcoupling is switched off, the clouds are projected back onto the atomic basis producing a second burst of atoms which may leave the trap region; however, a significant fraction of the atoms remain trapped. Hence a clean, quiet beam is not produced, and the bound dressed state is the cause of atom laser shutdown for strong coupling. Fig.\,\ref{Z-Raman_abs}(b) clearly demonstrates these features of atom laser shutdown for a Raman outcoupler operating between the Zeeman levels of the $F=1$ ground state of $^{87}$Rb (see Fig.\,\ref{Levels}(c)). The data is an integration of images taken for continuous output-coupling over 14ms as the coupling strength is increased.  The absorption image of each individual run of the experiment has been integrated in the direction perpendicular to propagation.  The figure is a plot of these integrated profiles as a function $\Omega_0$, which is the oscillation frequency of the untrapped state population for a semiclassical Rabi-flopping model. A continuous atom laser beam can clearly be seen at low frequencies, with the onset of complex outcoupling dynamics, and shutdown, as the coupling strength is increased. Fig.\,2(a) is a selection of data used to calibrate the Rabi frequency, and hence $\Omega_0$; the details of this calibration technique can be found in Section III. One should note that although for two-state coupling $\Omega_0=\Omega/2\pi$, in the case of three-state coupling, $\Omega_0=2^{3/2}\Omega/2\pi$.

We define the weak outcoupling limit as coupling strengths corresponding to an irreversible process that couples trapped atoms to a continuum of free falling states. This irreversibility is due to gravity (or any additional momentum transfer when using a Raman outcoupler) removing atoms from the coupling region. As an irreversible process, one may use Fermi's golden rule to calculate the the output flux, and hence for weak outcoupling, the flux is proportional to $\Omega_0^2$. The strong outcoupling limit can then be defined as coupling strengths corresponding to a reversible process, in which the external degrees of freedom can be neglected such that a semiclassical Rabi-flopping model is appropriate. The boundary between these two limits corresponds to the onset of complex outcoupling dynamics and atom laser shutdown. This intermediate region can be estimated, following the model in \cite{Robins:2006aa}, by comparing the timescales associated with $\Omega_0$ and that associated with the fall time $\tau_{fall}$ through the coupling region due to gravity and any momentum transfer due to the possible use of a Raman transition. As the coupling strength and hence $\Omega_0$ is increased, the time required for an oscillation of the untrapped state population becomes comparable to or less than the fall time, and one can no longer consider the effect of gravity to be irreversible. Some atoms can be coupled back into the condensate state and remain reasonably localized within the coupling region. It is by this reasoning that $\Omega_0$, and not $\Omega$ has been used as the parameter for comparison of RF and Raman based outcouplers. Clearly any momentum imparted by a Raman transition will reduce $\tau_{fall}$, enabling a larger value of $\Omega_0$ to be used before reaching the boundary between the strong and weak outcoupling regimes; hence a Raman outcoupler will result in a larger flux than an RF outcoupler, whilst still remaining in the weak outcoupling regime.

These features can be equivalently understood using the dressed state formalism. The dressed states are true eigenstates only in the limit that the coupling energy term in the Hamiltonian is infinitely larger than the kinetic term. In fact, the kinetic energy term in the Hamiltonian leads to off diagonal terms in the dressed basis, and therefore coupling of the dressed states.  In the work of Zobay \emph{et al.} \cite{Zobay:2004aa}, it is shown that the decay rate of atoms from the $|+\rangle$ state is given by $\gamma \sim 2\exp[-\pi\frac{\Omega^{3/2}}{\sqrt{2}\Delta^{1/2}\omega_z}]$. Thus the strong coupling condition is $\Omega\gg(2\omega_z^2\Delta/\pi^2)^{1/3}$, and as $\Delta$ represents the gravitational sag (hence the strength of gravity), this is in qualitative agreement with the semiclassical model used above. By transferring momentum to the outcoupled atoms, an additional kinetic energy term is added to the Hamiltonian, therefore increasing the Rabi frequency at which coupling between dressed states becomes negligible. Thus in the limit of weak coupling, the kinetic term must be included and the wave packet describing the BEC readily leaves the crossing region forming a typical atom laser beam. In the weak limit, the dressed basis treatment is therefore unnecessary, and one again employes Fermi's golden rule to calculate the flux. The fundamental principles derived in this section for a two-level system are easily translated to multi-level systems. 

\section{The Experiment}

\noindent Our production of BEC is discussed in detail elsewhere \cite{Doring:2008aa}. Briefly, condensates of approximately $2\times10^5$ atoms of $^{87}$Rb are prepared in the $|F=1,m_F=-1\rangle$ state in a magnetic Ioffe-Pritchard trap, with radial and longitudinal trapping frequencies of $\omega_z=2\pi\times120$\,Hz and $\omega_y=2\pi\times13$\,Hz respectively. To stabilize our trap we use low-noise power supplies and temperature controlled water cooling, resulting in a highly stable trap minimum that allows us to reproducibly address the condensate for atom laser production. Radio frequency outcoupling is performed using an rf loop antenna driven directly by a signal generator to couple the three Zeeman states in the $F=1$ manifold. The details of our two Raman outcouplers are described in detail for Zeeman coupling in \cite{Dugue:2008aa}, and for hyperfine coupling in \cite{Debs:2009aa}. Briefly, for hyperfine-Raman outcoupling we drive a two-photon transition between $|F=1,m_F=-1\rangle$ and $|F=2,m_F=0\rangle$ using two phase-locked optical beams separated in frequency by approximately $6.834$\,GHz, and detuned from the $5^2P_{3/2}$ resonance by $\Delta_R\simeq$ 90\,GHz. The Raman beams are produced by sourcing two beams from a single diode laser, and sending one of these through an electro-optic phase modulator, driven by a signal generator at approximately $6.834$\,GHz. The beams are directed onto the condensate orthogonal to one another and with appropriate polarization, as shown in Fig.\,\ref{Levels}(a). This results in a momentum kick of $\sqrt{2}\hbar k$ at 45$^\circ$ to the direction of gravity.

For Zeeman-Raman outcoupling we drive a two-photon transition between $|F=1,m_F=-1\rangle$ and $|F=1,m_F=0\rangle$ using two phase locked optical beams separated in frequency by approximately $1.34$\,MHz, and detuned from the $5^2P_{3/2}$ resonance by $\Delta_R\simeq$ 300\,GHz. The Raman beams are produced by sourcing two beams from a single diode laser, sending each of these beams through an acousto-optic modulator (AOM) in a double pass configuration. Each modulator is driven by phase locked signal generators, separated in frequency by approximately 0.67\,MHz. The beams are directed onto the condensate  as shown in Fig.\,\ref{Levels}(b). They are co-planar with gravity and the magnetic trap bias field,  separated by $\theta=140^{\circ}$ and given appropriate polarization to optimize the $\Delta m_F=1$ transition (see Fig.\,\ref{Levels}(c)). This results in a momentum kick of $2\hbar k\sin{\theta/2}\simeq 1.8\hbar k$ parallel to gravity. All three outcouplers are operated on resonance, as depicted in Fig.\,\ref{Levels}(c).

\begin{figure}[t]
\includegraphics[scale=.5]{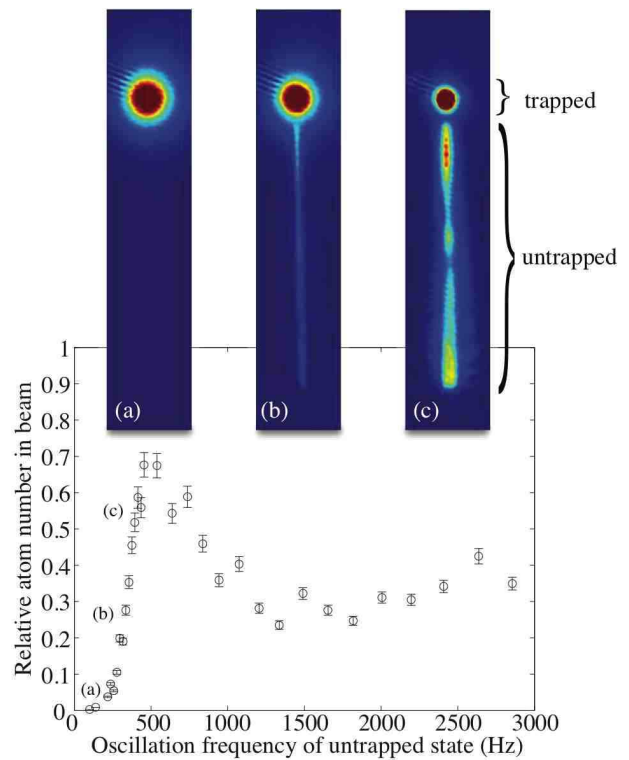}
\caption{\label{Z-Raman_bound} (Colour online) Relative number of atoms in the atom laser beam for Zeeman-Raman outcoupling as a function of $\Omega_0$. At $\Omega_0=500$\,Hz the number of atoms in the atom laser begins to decrease showing a clear effect of the bound dressed state shutting down the operation of the outcoupler. Absorption images are shown for each of the specified data points, and correspond to 14 ms of outcoupling as described in Section III. Error bars represent statistical uncertainty in the total number of atoms.}
\end{figure}

The data used to demonstrate the atom laser shutdown (Fig.\,\ref{Z-Raman_abs}(b)) were taken using the following outcoupling scheme. Raman coupling between Zeeman levels is switched on suddenly ($<200$\,ns) projecting the condensate onto the dressed basis (see Section II). The coupling field remains on for 14\,ms and is then suddenly switched off projecting back onto the bare atomic basis. The system is left to evolve for 5\,ms before the trap is switched off, and for a further 2\,ms before standard absorption imaging along the weak trapping direction ($y$ in Fig.\ref{Levels}(a)). The relative number of atoms transferred to the untrapped (atom laser) state is measured and plotted as a function of $\Omega_0$ in Fig.\,\ref{Z-Raman_bound}. These data show the dramatic effect of atom laser shutdown. For low $\Omega_0$, a continuous and clean beam is extracted. At $\Omega_0=500$\,Hz the effect of the bound dressed state manifests, causing an increasing number of atoms to remain trapped. As the coupling strength is further increased, the fraction of untrapped atoms saturates at around 0.35, dictated by the projection onto and from the dressed basis.

The Rabi frequency for each of our systems is calibrated using a method described in detail in \cite{Debs:2009aa}. Briefly, an outcoupling pulse of $100\,\mu$s) is applied to the trapped cloud, outcoupling a pulse of atoms. The relative number of atoms transferred to each state is measured as a function of the rf voltage or Raman beam power. These data are then fitted by a numerical simulation of the Gross-Pitaevski equation for the given system and with free parameters, allowing the Rabi frequency to be extracted as a function of voltage for rf, and beam power for Raman outcoupling. Fig.\,2(a) contains a selection of the images used to calibrate the Rabi frequency for the Zeeman-Raman outcoupler. 

\section{Comparison of Raman and rf outcouplers}

\begin{figure}[b]
\includegraphics{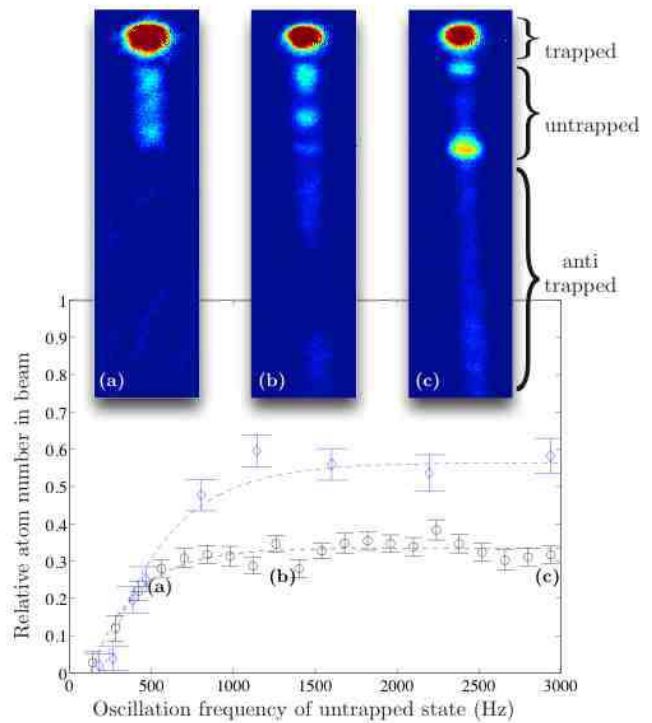}
\caption{\label{bound} (Colour online) Relative atom number in the atom laser beam as a function of $\Omega_0$. Black circles correspond to rf data, and blue diamonds to the Raman data. The dashed lines are fits to the data of the form $y=A(1-e^{-(x-x_0)/r})$, and allow a comparison of the bound state onset for each data set via the free parameter, $r$. Absorption images are shown for each of the specified points in the rf data. Error bars represent uncertainty in the fitted atom number for each image.}
\end{figure}

\noindent We now present a comparison of the rf and (two-state) hyperfine-Raman outcouplers. For these data, several changes are made to the outcoupling sequence given in Section III. Firstly, the coupling field remains on for only 3\,ms in order to allow imaging all atoms (trapped, untrapped, and anti-trapped). For the rf outcoupler, the system is left to evolve for 800\,$\mu$s after the coupling is switched off. This evolution time was maximized in order to separate the three magnetic sub-states as much as possible while still imaging all atoms onto the CCD camera. For the Raman outcoupler, the system can be evolved for 3.5\,ms after the coupling is switched off due to the absence of any anti-trapped states. The clouds are left to expand for 4.5\,ms after the trap switch off and a standard absorption image is taken along the radial trapping direction ($x$ in Fig.\ref{Levels}(a)).  The sequence is repeated for different coupling strengths and the relative number of atoms transferred to the untrapped state is plotted as a function of $\Omega_0$ in Fig.\,\ref{bound} for rf (black circles) and Raman (blue diamonds) outcoupling. 

Three examples of the absorption images used to extract the atom number for rf data are shown in Fig.\,\ref{bound}, as well as regions of interest corresponding to trapped, untrapped, and anti-trapped atoms. Atom numbers are extracted by integrating a Gaussian fitted to each row of an image and summing over all rows for a given region. To a good approximation, the `trapped,' `untrapped,' and `anti-trapped' regions correspond to $m_F=-1,\,0,\,$and 1 respectively, and the expected features are visible in all three images. For weak coupling (Fig.\,\ref{bound}(a)), a reasonably clean beam is seen, 3\,ms in length. This defines the region in which untrapped atoms will lie for all images. There are no discernible anti-trapped atoms. At the other extreme (Fig.\,\ref{bound}(c)), the strong coupling image shows a burst of atoms at the bottom of the `untrapped' region, corresponding to the initial projection, as well as a second burst at the top of the `untrapped' region corresponding to the final projection. A cloud of accelerated atoms defines the `anti-trapped' region, corresponding to atoms coupled to the anti-trapped state. Finally, as was the case for Fig.\,\ref{Z-Raman_abs}(b), for intermediate coupling strength (Fig.\,\ref{bound}(b)), complex dynamics are seen in the form of spatial oscillations in the atomic density. It should be noted that as $m_F=0$ atoms evolve freely under gravity (neglecting second order Zeeman effects), it is likely that upon projecting back to the atomic basis for the strongly coupled system, some of them will mix in with other states in our defined regions (e.g. in the anti-trapped region). However, our fundamental interest is in the onset of the bound state and atom laser shutdown, thus the finer details of the strong coupling regime have not been investigated in this work.

In the graph of Fig.\,\ref{bound}, the behavior theoretically predicted in section II is observed. For weak outcoupling ($\Omega_0<500$\,Hz), both data sets coincide. More importantly, as coupling strength is increased, a plateau is seen in the number of atoms transferred to the untrapped state for both rf and Raman outcoupling. This plateau occurs at a lower value of $\Omega_0$ for rf outcoupling, and also saturates at a smaller fraction of atoms outcoupled. The dashed lines are fits to the data of the form $y=A(1-e^{-(x-x_0)/r})$, and the parameter $r$ is used to compare the maximum $\Omega_0$ for the two systems. We find $\Omega_0$ is larger by a factor of 1.45 for Raman outcoupling, translating to an increase in flux of approximately 2.1. By utilizing a full 2$\hbar k$ momentum kick, this could be improved to an overall increase in flux by a factor of 5 for (two-state) Raman outcoupling over rf outcoupling. It is worth noting that when considering the two-state Raman outcoupler, the semiclassical model of section II predicts a maximum $\Omega_0$ of $\sim 1$\,kHz, which is in reasonable agreement with the experimental data. 

A final important result is that the three-state Raman outcoupler data in Fig.\,\ref{Z-Raman_bound} plateaus at a higher relative number than the rf outcoupler, but at a lower  relative number than the two-state Raman coupler. This provides experimental evidence that a two-state outcoupler will achieve a higher maximum flux than a three-state outcoupler, as predicted in \cite{Dugue:2007aa}. Thus, by combining Figs. \ref{Z-Raman_bound} and \ref{bound}, we find that a two-state Raman outcoupler can produce the highest continuous atom laser flux of any outcoupler for magnetically-confined samples. 

\section{conclusion}
\noindent When momentum is imparted to atoms outcoupled from a BEC to form an atom laser beam, a higher continuous flux is achievable compared with zero momentum transfer systems. We have experimentally verified that a two-state Raman outcoupler can achieve higher continuous flux than any rf-based or multi-level system. Coupled with the previous work on divergence and the spatial mode of Raman outcoupled beams, it is now clear that a two-level Raman outcoupler produces the highest-brightness atom laser beam of any outcoupler to date for magnetically confined samples. Furthermore, this work has shown that by using larger ($n2\hbar k$) momentum transfer during outcoupling, one can dramatically boost the flux of an atom laser beam. We are currently investigating several techniques for achieving high momentum transfer with a two-state Raman outcoupler, which will also enable large momentum transfer atomic beamsplitters for atom lasers. With rapidly developing technology for producing larger condensates with shorter machine duty cycles, the recent work on pumping \cite{Robins:2008aa,Doring:2009aa} and development of an atom laser Ramsey interferometer \cite{Doring:2008ab}, as well as the potential of utilizing squeezing, the atom laser is becoming a strong contender as a beam source for applications of atom interferometry. 

\acknowledgments
This work was supported by the Centre of Excellence program of the Australian Research Council, and the Alexander von Humboldt Foundation. JTS is thankful for a Fulbright Post-graduate Fellowship. 

\bibstyle{apsrev}
\bibliography{bib_min}

\begin{thebibliography}{50}
\bibitem[{fno({\natexlab{a}})}]{fnote1}
\bibinfo{howpublished}{Technical noise here refers to all forms of noise
  arrising from coupling to the environment. For magnetically confined BECs,
  any magnetic noise can couple to the condensate, including mechanical noice
  of the trap mount itself. Guided beams can also suffer due to mechanical
  noise. Although one can produce reasonably stable magnetic and mechanical
  systems, magnetically insensitive atom laser beams are far less susceptible
  to such effects.}
\bibitem[{fno({\natexlab{b}})}]{fnote2}
\bibinfo{howpublished}{Although it is true that the accumulated phase shift
  increases with area for both rotation and acceleration, the ratio of phase
  shifts due to rotation and acceleration increases with the aspect ratio of
  the enclosed area for an interferometer of given length. Hence, if one is
  interested in measuring rotations, a long and thin enclosed area is
  technically useful for minimizing the effects of acceleration noise, and a
  faster beam is desirable. The opposite is true for measuring accelerations.
  Of course, one requires a sufficient area to achieve the required
  sensitivity, and thus a compromise must be reached for a single beam device.
  This said, state of the art devices run two counter-propagating atom
  interferometers simultaneously, which allows cancellation of either the
  rotation or acceleration signal by subtracting or adding the output of each
  interferometer respectively. Provided the devices are well overlapped, and
  have low technical noise, this renders slow beams more useful, in general,
  for high sensitivity devices.}
\end{thebibliography}

\end{document}